\documentclass[letterpaper, 10 pt, conference]{ieeeconf}

\IEEEoverridecommandlockouts                              

\usepackage{tikz}
\usepackage{textcomp}

\usetikzlibrary{matrix}
\usetikzlibrary{angles,quotes}

\usepackage{graphics,graphicx} 
\usepackage{epsfig} 
\usepackage{mathptmx} 
\usepackage{times} 
\usepackage{amsmath} 
\usepackage{amssymb,nth, mathtools,dsfont}  
\usepackage{amsbsy}
\usepackage{epstopdf}
\usepackage[noadjust]{cite}
\usepackage{romannum}
\usepackage{xcolor}
\usepackage{subcaption}
\usepackage{multirow}
\usepackage{comment}

\newcommand{\bmtx}{\begin{bmatrix}}
\newcommand{\emtx}{\end{bmatrix}}
\newcommand{\bsmtx}{\left[ \begin{smallmatrix}} 
\newcommand{\esmtx}{\end{smallmatrix} \right]}
\newcommand{\field}[1]{\mathbb{#1}}
\newcommand{\R}{\field{R}}
\newcommand{\RH}{\field{RH}_\infty}

\newcommand{\N}{\field{N}}

\newcommand{\pjs}[1]{{\color{red}{(PJS: #1)}}}
\newcommand{\svn}[1]{{\color{blue}{(SVN: #1)}}}

\newtheorem{theorem}{Theorem}

\newtheorem{lemma}{Lemma}

\newtheorem{defin}{Definition}

\title{Discrete-Time Stability Analysis of ReLU Feedback \\ Systems via Integral Quadratic Constraints}

\author{Sahel Vahedi Noori, Bin Hu, Geir Dullerud, and Peter Seiler
	\thanks{S. Vahedi Noori and P. Seiler are with the Department of Electrical Engineering \& Computer Science, at the University of Michigan ({\tt\small sahelvn@umich.edu} and
		{\tt\small pseiler@umich.edu}). B. Hu is with the Department of Electrical and Computer Engineering at the University of Illinois at Urbana-Champaign ({\tt \small binhu7@illinois.edu}). G. Dullerud is with the Department of Mechanical Science and Engineering at the University of Illinois at Urbana-Champaign ({\tt \small dullerud@illinois.edu})
  }
	}

\begin{document}

\maketitle

\begin{abstract}
This paper analyzes internal stability of a discrete-time feedback system with a ReLU nonlinearity. This feedback system is motivated by recurrent neural networks. We first review existing static quadratic constraints (QCs) for slope-restricted nonlinearities. Next, we derive hard integral quadratic constraints (IQCs) for scalar ReLU by using finite impulse filters and structured matrices. These IQCs are combined with a dissipation inequality leading to an LMI condition that certifies internal stability. We show that our new dynamic IQCs for ReLU are a superset of the well-known Zames-Falb IQCs specified for slope-restricted nonlinearities. Numerical results show that the proposed hard IQCs give less conservative stability margins than Zames–Falb multipliers and prior static QC methods, sometimes dramatically so.

\end{abstract}


\section{Introduction}

Recurrent neural networks (RNNs) are widely used in machine learning for sequential data, with applications in speech, language, and time-series forecasting \cite{lecun2015deep, Goodfellow}. Their recurrent structure captures temporal dependencies, making them popular in reinforcement learning and control \cite{hochreiter1997long, pascanu2013difficulty}. However, this feedback structure can lead to exploding/vanishing gradients \cite{pascanu2013difficulty} and instability in practice. 

From a control-theoretic view, RNNs are Lurye-type feedback systems, with nonlinear activations wrapped around linear dynamics \cite{revay2020contracting, richardson23, hedesh2025robust}. Stability in such systems is essential for safety, robustness, and generalization. This motivates the use of tools like integral quadratic constraints (IQCs) \cite{megretski97} to provide performance guarantees for RNNs with ReLU activations. These tools vary in scope and formulation as summarized in Table~\ref{tab:LitReview}. Details of these methods are discussed in the following paragraphs.
\begin{table}[h!]
\centering
\scalebox{1}{
\begin{tabular}{ |c|c|c|c| } 
\hline
Method & Nonlinearity & Type & References\\ \hline
Static QCs & Slope-restricted & Repeated & \cite{Willems:71, Willems:1968} \\ \hline
Static QCs & ReLU & Repeated & \cite{noori24ReLURNN, noori25CompleteReLU, richardson23, ebihara2021stability, ebihara21}\\ \hline
Dynamic IQCs & Slope-restricted & Scalar & \cite{megretski97, Carrasco:2016, fetzer17, carrasco19}\\ \hline
Dynamic IQCs & ReLU & Scalar & Our Paper\\ \hline
Dynamic IQCs & Slope-restricted & Repeated & \cite{safonov2000zames, kulkarni2002all} \\ \hline
Dynamic IQCs & ReLU & Repeated & Future Work \\ \hline
\end{tabular}
}
\caption{Summary of prior methods for stability analysis of nonlinear feedback interconnections.}
    \label{tab:LitReview}
\end{table}

Static quadratic constraints (QCs) provide tractable conditions for stability analysis of systems with slope-restricted nonlinearities. For functions with slope in $[0,1]$ and $\phi(0)=0$, pointwise inequalities can be derived and, through lifting, extended to repeated nonlinearities. These results, established in classical works such as \cite{Willems:71, Willems:1968}, form the basis for time-domain stability analysis.

Static QCs have also been derived specifically for the ReLU function. In previous works \cite{noori24ReLURNN, noori25CompleteReLU, richardson23, ebihara2021stability, ebihara21}, pointwise QCs for the repeated ReLU are obtained, enabling stability and performance analysis of ReLU RNNs. By exploiting ReLU properties, these results yield tighter guarantees than generic slope-restricted QCs.

Dynamic IQCs can also be specified in the time domain or frequency domain for slope-restricted nonlinearities. A key tool is the Zames-Falb multiplier \cite{megretski97, fetzer17, carrasco19}. Stability can be certified through a frequency domain inequality with an appropriate multiplier.  This generalizes the classical circle and Popov. This leads to the more general IQC framework, introduced by Megretski and Rantzer \cite{megretski97}. Follow-up works developed the framework in several directions. Convex searches over FIR multipliers have been proposed \cite{Carrasco:2016}. The framework has also been extended to discrete-time systems with both causal and anticausal multipliers \cite{fetzer17csl, carrasco19}. In addition, less conservative LMI conditions have been introduced \cite{ahmad2014less}.

In this paper, we develop a discrete-time stability analysis framework for ReLU feedback systems using hard IQCs. Our main contribution lies in deriving a new dynamic IQC for scalar ReLU and proving that this class contains all Zames–Falb multipliers for $[0,1]$ slope-restricted nonlinearities. Numerical comparisons with the FIR multiplier-based method from \cite{carrasco19} and our previous work \cite{noori24ReLURNN}, show that our new approach achieves improved performance for ReLU nonlinearities. Our work fills the open row in Table~\ref{tab:LitReview} for scalar ReLU. It also opens the door, as future work, to develop dynamic IQCs for repeated ReLU.

\section{Notation}

This section briefly reviews basic notation regarding vectors, matrices, and signals.  Let $\R^n$ and $\R^{n\times m}$ denote the sets of real $n\times 1$
vectors and $n\times m$ matrices, respectively. Moreover, $\R_{\ge 0}^n$ and $\R_{\ge 0}^{n\times m}$ denote vectors and matrices of the given
dimensions with non-negative entries.   $M\in \R^{n\times n}$ is a \emph{Metzler matrix} if the off-diagonal entries are non-negative, i.e. $M_{ij}\ge 0$ for $i\ne j$.  A matrix $M\in \R^{n\times n}$ is \emph{doubly hyperdominant} if the off-diagonal elements are non-positive, and both the row sums and column sums are non-negative.  Finally, a matrix $M \in \R^{n\times m}$ is \emph{Toeplitz} if it is constant along each diagonal, i.e., $A_{i,j} = A_{i+k,j+k}$ for any integers $(i,j,k)$ such that $1 \le i, i+k \le n$ and $1 \le j, j+k \le m$.

Next, let $\N$ denote the set of non-negative integers.  Let $v:\N \to \R^n$ and $w:\N \to \R^n$ be real, vector-valued sequences.  Define the inner product $\langle v,w \rangle : = \sum_{k=0}^\infty v(k)^\top w(k)$.
A sequence $v$ is said to be in $\ell_2^n$ if $\langle v,v\rangle <\infty$. In addition, the $2$-norm for $v \in \ell_2^n$ is defined as
$\|v\|:=\sqrt{ \langle v,v\rangle}$.   We'll use $\ell_{2e}$ to denote the extended space of sequences whose $\ell_2$ norm is not necessarily finite.
$\mathbb{RL_\infty}$ denotes the set of rational functions with real coefficients that have no poles on the unit circle and $\mathbb{RH_\infty}$ is the subset of functions in $\mathbb{RL_\infty}$ with all poles inside the unit disk of the complex plane.


\section{Problem statement}

\begin{figure}[h!t]
\centering
\begin{picture}(230,90)(40,20)
 \thicklines
 \put(80,40){\framebox(30,30){$G$}}
 \put(80,75){\framebox(30,30){$\phi$}}
 \put(36,70){$v(k)$}
 \put(55,55){\line(1,0){25}}  
 \put(55,55){\line(0,1){35}}  
 \put(55,90){\vector(1,0){25}}  
 \put(138,70){$w(k)$}
 \put(135,90){\line(-1,0){25}}  
 \put(135,55){\line(0,1){35}}  
 \put(135,55){\vector(-1,0){25}}  
 \put(170,45){\vector(1,0){100}}  
 \put(260,50){$v(k)$}
 \put(220,25){\vector(0,1){70}}  
 \put(191,100){$w(k)=\phi(v(k))$}
 \put(170,46){{\color{red} \line(1,0){50}} }  
 \put(220,46){{\color{red} \line(1,1){35}} }  
\end{picture}
\caption{Left: Interconnection $F_U(G,\phi)$ of a nominal discrete-time LTI system $G$ and scalar ReLU $\phi$. Right: Graph of scalar ReLU $\phi$.}
\label{fig:LFTdiagram}
\end{figure}
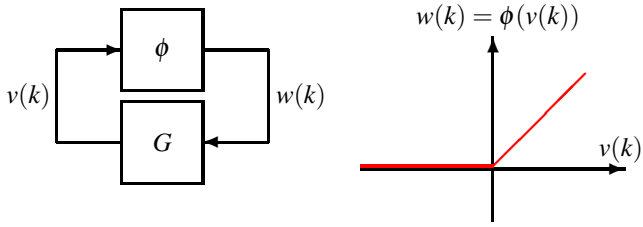

Consider the interconnection shown on the left of Figure~\ref{fig:LFTdiagram}  with a static nonlinearity $\phi$ wrapped in feedback around the channels
of a nominal system $G$.  This interconnection is denoted as $F_U(G,\phi)$.  

The nominal part $G$ is a discrete-time, linear time-invariant (LTI)
system described by the following state-space model:
\begin{align}
  \label{eq:LTInom}
  \begin{split}
    x(k+1) & = A\, x(k) + B\,w(k) \\
    v(k) & = C\,x(k)+D\, w(k),
  \end{split}
\end{align}
where $x(k) \in \R^{n_x}$, $w(k)\in \R$, and $v(k)\in \R$ are the state, input and output at time $k$.  The static nonlinearity $\phi:\R \to \R_{\ge 0}$ is the ReLU function. The ReLU, shown on the right of
Figure~\ref{fig:LFTdiagram}, is:
\begin{align}
\label{eq:ReLU}
\phi(v(k))= \left\{
  \begin{array}{ll}
    0 & \mbox{if } v(k) < 0 \\
    v(k) & \mbox{if } v(k) \geq 0 
  \end{array} 
  \right. .
\end{align}
The interconnection $F_U(G,\phi)$ is known as a Lurye decomposition in the robust control literature \cite{khalil01}.

This feedback interconnection involves an implicit equation if $D\ne 0$.  Specifically, the second equation in \eqref{eq:LTInom} combined with $w(k)=\phi( v(k) )$ yields:
\begin{align}
   \label{eq:WellPosed}
   v(k)=C\,x(k)+D\, \phi( v(k) ).
\end{align}
This equation is \emph{well-posed} if there exists a unique solution $v(k)$ for all values of $x(k)$.  Well-posedness of this equation implies that the dynamic system $F_U(G,\phi)$ is well-posed in the following sense:  for all initial conditions $x(0)\in\R^{n_x}$ there exists unique signals $x$, $v$, $w$
that satisfy the interconnection defined by \eqref{eq:LTInom} and  \eqref{eq:ReLU}.
There are simple sufficient conditions for well-posedness of \eqref{eq:WellPosed}, e.g. Lemma 1 in \cite{richardson23} (which relies on results in \cite{valmorbida18,zaccarian02}). Thus, we will assume well-posedness for simplicity in the remainder of the paper.

A well-posed interconnection $F_U(G,\phi)$ is \emph{internally stable} if the state trajectory from any initial condition $x(0)$ satisfies $x(k)\to 0$ as $k\to \infty$. In other words, $F_U(G,\phi)$ is internally stable if the origin is a globally asymptotically stable equilibrium point for the interconnection defined by \eqref{eq:LTInom} and  \eqref{eq:ReLU}. The remainder of the paper will focus on sufficient conditions for stability of the interconnection $F_U(G,\phi)$. The stability conditions will combine Lyapunov theory and integral quadratic constraints for ReLU. Our main contribution will be a novel dynamic IQC derived specifically for the ReLU function.



\section{Preliminary Results}

In this section, we summarize existing results for a nonlinearity $\phi:\R \to \R_{\ge 0}$ that satisfies $\phi(0)=0$ and has slope restricted to $[0,1]$.  The $[0,1]$ slope restriction corresponds to the following constraint:
\begin{align}
  0 \le  \frac{ \phi(v_2)-\phi(v_1)}{v_2-v_1} \le 1, 
  \,\, \forall v_1, \, v_2 \in \R, \, v_1\ne v_2.
\end{align}
The class of $[0,1]$ slope restricted functions includes ReLU as a special case.  The rest of this section describes known static QCs and dynamic IQCs for slope restricted functions. We also state an internal stability condition that incorporates the dynamic IQC.  These existing results form the basis for a new dynamic IQC for ReLU as derived in the next section.

\subsection{Static Quadratic Constraints}
\label{sec:SlopeResQCs}

In this subsection we review an existing static QC for scalar functions $\phi$ with slope restricted to $[0,1]$. First, consider the nonlinearity $\Phi:\R^{N} \to \R^{N}$ defined by element-wise application of the scalar nonlinearity:
\begin{align}
\label{eq:Phi}
	\Phi(v) := \bmtx \phi(v_1), \, \phi(v_2), \, \ldots, \phi(v_{N}) \emtx^\top.
\end{align}
The next lemma presents a static QC for this class of repeated slope-restricted functions.
\vspace{0.1in}
\begin{lemma}
\label{lem:doublyhypQC}
Let $\Phi:\R^{N} \to \R^{N}$ be a repeated nonlinearity defined element-wise by $\phi: \R \to \R$ where $\phi(0)=0$ and $\phi$ is slope restricted to $[0,1]$. Moreover, let $Q_0 \in \R^{N \times N}$ be any doubly hyperdominant matrix
and define
\begin{align}
    Q:= \bmtx 0 & Q_0^\top \\ Q_0 & -(Q_0+Q_0^\top) \emtx,
\end{align}
Then the following QC holds  $\forall v\in\R^{N}$ and $w=\Phi(v)$:
\begin{align}
\label{eq:doublyhypQC}
    \bmtx v \\ w \emtx^\top
     Q   
    \bmtx v \\ w \emtx \ge 0.
\end{align}
\end{lemma}
\vspace{0.1in}
\begin{proof}
This follows from the results in Section 3.5 of \cite{Willems:71}.
\end{proof}
\vspace{0.1in}

\subsection{Dynamic Integral Quadratic Constraints}
\label{sec:SlopeResIQCs}

In this subsection, we review a dynamic IQC for scalar functions $\phi:\R \to \R$ where $\phi(0)=0$ and with slope restricted to $[0,1]$. This is a version of the Zames-Falb multiplier \cite{Carrasco:2016,fetzer17,carrasco19, Willems:1968}.

Consider a sequence $v\in\ell_{2e}$ and define a new sequence $w\in \ell_{2e}$ by applying this function at each time, i.e., $w(k)=\phi(v(k))$ for all $k\in \N$.  With a slight abuse of notation, we use $\phi:\ell_{2e} \to \ell_{2e}$ to also denote this nonlinearity acting on sequences.  A dynamic IQC can be constructed for $\phi$ based on the static QC introduced in the previous section for the repeated nonlinearity $\Phi$. The formal definition for a dynamic IQC is provided next.

\vspace{0.1in}
\begin{defin}
  \label{def:tdiqc}
  Let $\Psi \in \RH^{n_r \times 2}$ and $M=M^\top \in \R^{n_r \times n_r}$ be given. A nonlinearity $\phi: \ell_{2e} \to \ell_{2e}$ satisfies the time domain, hard IQC defined by $(\Psi, M)$ if the following inequality holds for all $v \in \ell_{2e}$, $w=\phi(v)$ and  for all $T_0\ge 0$,
    \begin{align}
      \label{eq:tdhardiqc}
      \sum_{k=0}^{T_0} r(k)^\top M r(k)  \ge 0,
    \end{align}
    where $r$ is the output of $\Psi$ driven by inputs $(v,w)$ with zero initial conditions.
\end{defin}
\vspace{0.1in}

This definition is called a \emph{hard} IQC because the constraint~\eqref{eq:tdhardiqc} holds for all $T_0\ge 0$.  
This is in contrast to \emph{soft} IQCs in the literature
which only hold as $T_0\to \infty$ (and require the additional assumption that $v$, $w\in \ell_2$ to ensure the summation in  \eqref{eq:tdhardiqc} remains finite).  

A hard IQC for slope-restricted functions can be defined using the following filter $\Psi_N(z)$:
\begin{align}
    \label{eq:FiniteFilter}
    \Psi_N(z) &:= \left[\begin{array}{cccc|cccc}
        1 & z^{-1}  & \cdots & z^{-N} & 0 & 0 & \cdots & 0\\
        \hline
        0 & 0 & \cdots & 0 & 1 & z^{-1} & \cdots & z^{-N}
        \end{array}\right]^\top.
\end{align}
Exciting this filter with an input sequence $\bsmtx v \\ w \esmtx$ from zero initial conditions gives the following output at time $k$:
\begin{align}
\label{eq:rN}
r(k) := \bmtx v(k), \ldots, v(k-N), \, w(k), \ldots, w(k-N) \emtx^\top \in \R^{2N+2}.
\end{align}
The assumption of zero initial conditions implies that $v(j)=0$ and $w(j)=0$ for all $j<0$. The next lemma states a dynamic IQC for $\phi$ using this filter. This is a known result (see Section IV.B of \cite{carrasco19}), but the proof is given as it provides the starting point for the main result in the next section.
\vspace{0.1in}
\begin{lemma}
\label{lem:hardIQCSlope}
Let $\phi:\ell_{2e} \to \ell_{2e}$ be an element-wise nonlinearity with $\phi(0)=0$ and slope restricted to $[0,1]$.   Moreover, let $\{m_i\}_{i=-N}^N \subset \R$ be given with  $m_i \le 0 $ for $i\ne 0$ and $\sum_{i=-N}^N m_i \ge 0$. Then $\phi$ satisfies the time-domain hard IQC defined by $(\Psi_N,M)$ where
\begin{align}
    \label{eq:M_SlopeRes}
    M := \bmtx 0 & M_0^\top \\ M_0 & -(M_0+M_0^\top) \emtx,
\end{align}
with
\begin{align}
    \label{eq:M0}
    M_0 := 
    \bmtx m_0 & m_1 & \cdots & m_N \\
    m_{-1} & 0 & \cdots & 0\\
    \vdots \\
    m_{-N} & 0 & \cdots & 0 \emtx.
\end{align}
\end{lemma}
\vspace{0.1in}
\begin{proof}
For any $T_0\ge 0$, define the following stacked vectors of the signals $v$ and $w=\phi(v)$ from $t=0$ through $t=T_0$,
\begin{align}
    \label{eq:stackvecs1}
    \begin{split}
        \bar{v}_{T_0} & := \bmtx v(T_0) & v(T_0-1) & \cdots & v(0) \emtx^\top \in \R^{T_0+1}, \\
        \bar{w}_{T_0} & := \bmtx w(T_0) & w(T_0-1) & \cdots & w(0) \emtx^\top \in \R^{T_0+1}.
    \end{split}
\end{align}
Next, define a Toeplitz matrix $Q_0 \in \R^{(T_0+1)\times (T_0+1)}$ with the first row and column
defined by
\begin{align*}
Q_0(1,:) & := \bmtx m_0 & m_{1} & \cdots & m_{N}
    & 0 & \cdots & 0 \emtx \in \R^{1 \times (T_0+1)}, \\
Q_0(:,1) & := \bmtx m_0 & m_{-1} & \cdots & m_{-N}
    & 0 & \cdots & 0 \emtx^\top \in \R^{(T_0+1) \times 1}. 
\end{align*}
If $T_0<N$ then the first row and column of $Q_0$ is defined using only the entries $\{m_i\}_{i=-T_0}^{T_0}$ (and the additional zeros will not appear).  In any case, the assumptions on $\{ m_i \}_{i=-N}^N$ imply that
$Q_0$ is a doubly hyperdominant matrix.

Finally, it can be shown, by direct substitution, that the
output $r$ of $\Psi_N$ driven by $(v,w)$ from zero
initial conditions satisfies the following for any
$T_0\ge 0$: 
\begin{align*}
  \sum_{k=0}^{T_0} r(k)^\top M r(k)  
  = \bmtx \bar{v}_{T_0}  \\ \bar{w}_{T_0} \emtx^\top
    \bmtx 0 & Q_0^\top \\ Q_0 & -(Q_0+Q_0^\top) \emtx
    \bmtx \bar{v}_{T_0}  \\ \bar{w}_{T_0} \emtx
\end{align*}
This is nonnegative for any $T_0\ge 0$ by
Lemma~\ref{lem:doublyhypQC}.
\end{proof}
\vspace{0.1in}

Lemma~\ref{lem:hardIQCSlope} provides a time-domain representation for the IQC.  We can connect this to a frequency-domain constraint when $v \in \ell_2$.  If $\phi$ is $[0,1]$ slope restricted then $w=\phi(v)$ is also in $\ell_2$.  It follows that the hard IQC \eqref{eq:tdhardiqc} will converge to a finite value in the limit as $T_0\to \infty$. Moreover, the limit will be non-negative by Lemma~\ref{lem:hardIQCSlope}.  Hence the hard IQC implies that the soft IQC also holds.  The soft IQC can be transformed into a frequency-domain IQC using Parseval's theorem. This yields the discrete-time Zames–Falb multiplier $\sum_{i=-N}^N m_i z^{-i}$ \cite{Carrasco:2016,fetzer17,carrasco19, Willems:1968}.

\subsection{Stability Condition}
\label{sec:SlopeResStab}
This subsection reviews a sufficient condition for
internal stability  based on the hard IQC given in the previous subsection. Let $G(z) = C(zI-A)^{-1}B + D$ denote the transfer function of the LTI system \eqref{eq:LTInom} and $\Psi_N(z)$ be the finite impulse response filter defined in \eqref{eq:FiniteFilter}. Define the augmented system $\hat{G}$ as
\begin{align}
\label{eq:Ghat}
\hat{G}(z)= \Psi_N(z) \bmtx G(z) \\ 1 \emtx \sim 
\left[\begin{array}{c|c} \hat{A} & \hat{B}\\ \hline \hat{C} & \hat{D} \end{array}\right].
\end{align}
Moreover, define the following matrix function using the matrices of $\hat{G}$
\begin{align}
\label{eq:LMI}
LMI(P,M) := 
\bmtx \hat{A}^\top P \hat{A}-P  & \hat{A}^\top P \hat{B} \\ 
  \hat{B}^\top P \hat{A} & \hat{B}^\top P \hat{B} \emtx + \bmtx \hat{C}^\top \\ \hat{D}^\top \emtx M \bmtx \hat{C} & \hat{D} \emtx.
\end{align}
The next theorem provides a sufficient condition for internal stability in terms of  the linear matrix inequality $LMI(P,M) \prec 0$.

\vspace{0.1in}
\begin{theorem}
    \label{thm:slopeRes}
    Consider the interconnection $F_U(G,\phi)$ where
    $G$ is the LTI system \eqref{eq:LTInom}.  $\phi: \ell_{2e} \to \ell_{2e}$ is a scalar nonlinearity with $\phi(0)=0$ and slope restricted to $[0,1]$.  Assume the interconnection is well-posed.
    Let $\{m_i\}_{i=-N}^N \subset \R$ be given with $m_i \le 0 $ for $i\ne 0$ and $\sum_{i=-N}^N m_i \ge 0$. 
    Then the interconnection $F_U(G,\phi)$ is internally
    stable if there exists $P\succeq 0$ such
    that $LMI(P,M) \prec 0$ where $M$ is defined 
    as in Equations~\ref{eq:M_SlopeRes} and \ref{eq:M0}.
\end{theorem}
\vspace{0.1in}
\begin{proof} 
By well-posedness, for any $x(0) \in \R^{n_x}$ there is a unique solution $(x,v,w)$ to the interconnection $F_U(G,\phi)$.  Moreover, the filter $\Psi_N$ driven by $(v,w)$ from zero initial conditions has a unique output $r$. Let $\hat{x}$ denote the corresponding state trajectory for the augmented system $\hat{G}$.

The LMI is strictly feasible and hence it remains feasible under small perturbations: for a sufficiently small $\epsilon>0$ the perturbed matrix $\hat{P}:= P+\epsilon I \succ 0$ satisfies 
\begin{align}
LMI(\hat{P},M) + \epsilon \bmtx I & 0 \\ 0 & 0 \emtx
\prec 0.
\end{align}
Define a storage function by $V\left(\hat{x}\right) := \hat{x}^\top \hat{P} \hat{x}$.  Left and right multiply the perturbed LMI by $[\hat{x}(k)^\top \; w(k)^\top]$ and its transpose.  The result, applying the augmented dynamics~\eqref{eq:Ghat}, gives the following:
\begin{align*}
    V\left(\hat{x}(k+1)\right)  - V\left(\hat{x}(k)\right) + \epsilon \hat{x}(k)^\top \hat{x}(k) + 
     r(k)^\top M r(k)
    \le 0
\end{align*}
Summing this inequality from $k = 0$ to an arbitrary time  $k = T_0$ yields:
\begin{align}
    \label{eq:dissipationSum}
    \begin{split}        
    & V\left(\hat{x}(T_0+1)\right)  
    - V\left(\hat{x}(0)\right) + \epsilon \sum_{k=0}^{T_0} \hat{x}(k)^\top \hat{x}(k) \\
    & +  \sum_{k=0}^{T_0} r(k)^\top M r(k)
    \le 0
    \end{split}    
\end{align}
The first term is nonnegative since $\hat{P}\succ 0$
and the last term is nonnegative by Lemma \ref{lem:hardIQCSlope}.  Hence this implies:
\begin{align}
    \label{eq:stateNormBound}
    \epsilon \sum_{k=0}^{T_0} \hat{x}(k)^\top \hat{x}(k)
    \le V\left(\hat{x}(0)\right) 
\end{align}
This bound holds for any $T_0 \ge 0$. Take the limit as $T_0\to \infty$ to conclude that $\| \hat{x}\|_2^2 \le \frac{1}{\epsilon} V(\hat{x}(0))$.  It follows that the $\ell_2$-norm of $\hat{x}$ is also bounded since $V(\hat{x}(0))$ is finite. This implies $\hat{x}(k) \to 0$ as $k\to \infty$ and hence $x(k) \to 0$ as well.
\end{proof}
\vspace{0.1in}

Megretski and Rantzer \cite{megretski97} introduced an IQC framework for robustness and stability analysis of feedback systems using soft (or hard) IQCs to describe the perturbation. Their stability theorem guarantees input–output stability but does not require the Lyapunov matrix $P$ to be positive semidefinite. This is in contrast to Theorem~\ref{thm:slopeRes} above which proves internal stability but requires $P$ to be positive semidefinite. It is notable that when the multiplier is either purely causal ($m_i=0$ for $i<0$) or purely anticausal ($m_i=0$ for $i>0$) then the constraint $P\succ 0$ is automatically satisfied, as established in Section IV of \cite{fetzer17csl}.

\section{Main Results}

In this section, we present sufficient conditions for the stability and performance of a ReLU feedback system using hard IQCs. Following the structure of the previous section, we first review the static QC for the repeated ReLU derived in our earlier work \cite{noori24ReLURNN}. Next, we formulate a new hard dynamic IQC for the scalar ReLU. Finally, we adapt the stability analysis to incorporate our new IQC for the scalar ReLU.

\subsection{Static Quadratic Constraints}
\label{sec:ReLUQCs}

In this subsection we review an existing static QC for ReLU. The function is defined in \eqref{eq:ReLU} and illustrated on the right-hand side of Figure~\ref{fig:LFTdiagram}.

\vspace{0.1in}
\begin{lemma}
    \label{lem:pointwiseQCs}
    Let $\Phi:\R^{N} \to \R^{N}$, defined in \eqref{eq:Phi}, be the repeated ReLU where $\phi: \R \to \R$ is the scalar ReLU.
    Let $Q_1=Q_1^\top$, $Q_2=Q_2^\top\in \R_{\ge 0}^{N \times N}$, and $Q_3\in \R^{N \times N}$ be given with $Q_3$ a Metzler matrix. Define
    \begin{align}
        Q:=  \bmtx Q_1 & -Q^\top_3-Q_1 \\ -Q_3-Q_1 & Q_1+Q_2+Q_3+Q_3^\top\emtx.
    \end{align}
    Then the following QC holds  $\forall v\in\R^{N}$ and $w=\Phi(v)$:
    \begin{align}
    \label{eq:repReLUFin}
    \bmtx v \\ w\emtx^\top Q
    \bmtx v \\ w\emtx \ge 0
    \end{align}
\end{lemma}
\vspace{0.1in}
\begin{proof}
This follows from the QC results for ReLU in  \cite{richardson23, noori24ReLURNN, ebihara21}.
\end{proof}
\vspace{0.1in}

This lemma contains a large class of available static QCs for repeated ReLU.  However, there are other QCs that can be specified for repeated ReLU, e.g. QCs using copositive matrices as in \cite{ebihara21}. In fact, the complete class of QCs for repeated ReLU can specified by $2^N$ copositivity conditions (Section III.A of \cite{noori25CompleteReLU}). 
 We will next derive a dynamic IQC for scalar ReLU using Lemma~\ref{lem:pointwiseQCs}.
 It will be left as future work to see if the complete class (or other subclasses) of static QCs for repeated ReLU can be used to derive dynamic IQCs for scalar ReLU.

\subsection{Dynamic Integral Quadratic Constraints}
\label{sec:ReLUIQCs}

In this subsection, we derive a dynamic IQC for scalar ReLU following the same procedure as in Subsection \ref{sec:SlopeResIQCs}.

Consider sequences $v\in\ell_{2e}$ and $w\in \ell_{2e}$ where $\phi:\ell_{2e} \to \ell_{2e}$, $w_k =\phi(v_k)$ for all $k\in \N$. The next lemma presents the new dynamic IQC for scalar ReLU.  The proof is based on the static QC given for repeated ReLU in Lemma~\ref{lem:pointwiseQCs}.

\vspace{0.1in}
\begin{lemma}
\label{lem:hardIQCReLU}
Let $\phi:\ell_{2e} \to \ell_{2e}$ be an element-wise ReLU.   Moreover, let $\{m^1_i\}_{i=0}^N, \{m^2_i\}_{i=0}^N \subset \R_{\ge 0}$, and $\{m^3_i\}_{i=-N}^N \subset \R$ be given with  $m^3_i \ge 0 $ for $i\ne 0$. Then $\phi$ satisfies the time-domain hard IQC defined by $(\Psi_N,M)$ where
\begin{align}
\label{eq:M_ReLU}
   M:= \bmtx M_1 & -M^\top_3-M_1 \\ -M_3-M_1 & M_1+M_2+M_3+M_3^\top\emtx,
\end{align}
with
\begin{align}
    \label{eq:M12}
    M_j := 
    \bmtx m^j_0 & m^j_1 & \cdots & m^j_N \\
    m^j_1 & 0 & \cdots & 0\\
    \vdots \\
    m^j_N & 0 & \cdots & 0 \emtx, \;\; j=1,2,
\end{align}
\begin{align}
    \label{eq:M3}
    M_3 := 
    \bmtx m^3_0 & m^3_1 & \cdots & m^3_N \\
    m^3_{-1} & 0 & \cdots & 0\\
    \vdots \\
    m^3_{-N} & 0 & \cdots & 0 \emtx.
\end{align}
\end{lemma}
\vspace{0.1in}
\begin{proof}
The proof is similar to the proof of Lemma \ref{lem:hardIQCSlope}. We define the following three Toeplitz matrices $Q_1=Q_1^\top$, $Q_2=Q_2^\top \in \R^{(T_0+1)\times(T_0+1)}$, $Q_3 \in \R^{(T_0+1)\times (T_0+1)}$ with the first rows and columns defined by:
\begin{align*}
 Q_j(1,:) & = Q_j(:,1)^\top \\
    & := \bmtx m^j_0 & m^j_{1} & \cdots & m^j_{N}
    & 0 & \cdots & 0 \emtx \in \R^{1 \times (T_0+1)} \\
\end{align*}
for $j=1,2$, and
\begin{align*}
Q_3(1,:) & := \bmtx m^3_0 & m^3_{1} & \cdots & m^3_{N}
    & 0 & \cdots & 0 \emtx \in \R^{1 \times (T_0+1)} \\
Q_3(:,1) & := \bmtx m^3_0 & m^3_{-1} & \cdots & m^3_{-N}
    & 0 & \cdots & 0 \emtx^\top \in \R^{(T_0+1) \times 1} 
\end{align*}
Again, if $T_0<N$ then the first row and columns of $Q_3$ is defined using only the entries $\{m_i^3\}_{i=-T_0}^{T_0}$ (and the additional zeros will not appear).  A similar statement holds for $Q_1$ and $Q_2$. In any case, the assumptions on $\{ m^1_i \}_{i=0}^N$ and $\{ m^2_i \}_{i=0}^N$ imply that $Q_1=Q^\top_1,Q_2=Q^\top_2 \in \R^{(T_0+1)\times (T_0+1)}_{\ge 0}$. Moreover, the assumptions on $\{ m^3_i \}_{i=-N}^N$ imply that $Q_3$ is a Metzler matrix.

Finally, it can be shown, by direct substitution, that the output $r$ of $\Psi_N$ driven by $(v,w)$ from zero initial conditions satisfies the following for any $T_0\ge 0$: 
\begin{align*}
  \sum_{k=0}^{T_0} r(k)^\top M r(k)  
  = \bmtx \bar{v}_{T_0}  \\ \bar{w}_{T_0} \emtx^\top
    Q
    \bmtx \bar{v}_{T_0}  \\ \bar{w}_{T_0} \emtx
\end{align*}
where 
\begin{align*}
    Q=  \bmtx Q_1 & -Q^\top_3-Q_1 \\ -Q_3-Q_1 & Q_1+Q_2+Q_3+Q_3^\top\emtx.
\end{align*}
By Lemma \ref{lem:pointwiseQCs}, the QC on the right-hand side is nonnegative for all $T_0 \ge 0$.
\end{proof}
\vspace{0.1in}

The class of hard IQCs obtained in Lemma \ref{lem:hardIQCReLU} includes the Zames–Falb IQCs.  Specifically, consider 
a feasible Zames-Falb IQC  defined by $\{m_i\}_{i=-N}^N$ with  $m_i \le 0 $ for $i\ne 0$ and $\sum_{i=-N}^N m_i \ge 0$. 
We can define an IQC for ReLU as in Lemma \ref{lem:hardIQCReLU} using the following definitions:
\begin{align*}
& m_i^1, m_i^2 =0 \;\; \mbox{ for } i=0,\ldots N \\
& m_i^3 = -m_i \;\; \mbox{ for } i=-N, \ldots, N
\end{align*}
In other words, our dynamic IQCs for ReLU form a superset of all dynamic IQCs for $[0,1]$ slope-restricted nonlinearities.

\subsection{Stability Condition}
\label{sec:ReLUStab}
This subsection derives a sufficient condition for internal stability of interconnection $F_U(G,\phi)$ with $\phi:\ell_{2e} \to \ell_{2e}$ being an element-wise ReLU, based on the hard IQC derived in the previous subsection. Define the transfer function $G(z)$, filter $\Psi_N(z)$, the augmented system $\hat{G}$, and the linear matrix inequality $LMI(P,M)$ as in Section~\ref{sec:SlopeResStab}.

\vspace{0.1in}
\begin{theorem}
    \label{thm:ReLU}
    Consider the interconnection $F_U(G,\phi)$ where
    $G$ is the LTI system \eqref{eq:LTInom}.  $\phi: \ell_{2e} \to \ell_{2e}$ is an element-wise ReLU.  Assume the interconnection is well-posed.
    Let $\{m^1_i\}_{i=0}^N, \{m^2_i\}_{i=0}^N \subset \R_{\ge 0}$, and $\{m^3_i\}_{i=-N}^N \subset \R$ be given with  $m^3_i \ge 0 $ for $i\ne 0$. 
    Then the interconnection $F_U(G,\phi)$ is internally
    stable if there exists $P\succeq 0$ such
    that $LMI(P,M) \prec 0$ where $M$ is defined 
    as in Equations~\ref{eq:M_ReLU}, \ref{eq:M12} and \ref{eq:M3}.
\end{theorem}
\vspace{0.1in}
\begin{proof}
The proof follows similar to that given for Theorem \ref{thm:slopeRes}.  The only change is that the $M$ defined in Equations~\ref{eq:M_ReLU}, \ref{eq:M12} and \ref{eq:M3} yields a hard IQC term defined by $(\Psi_N, M)$ (Lemma \ref{lem:hardIQCReLU}). This hard IQC is used in the dissipation inequality to show $x(k) \to 0$ as $k \to \infty$.
\end{proof}

\section{Numerical examples}

A variety of examples are given in  \cite{carrasco19} to study the use of discrete-time Zames-Falb multipliers for analyzing stability of Lurye systems.  We previously used Example 6 in Table 1 of \cite{carrasco19} to compare with our results in \cite{noori24ReLURNN}.  We reexamine this example to illustrate our new IQC for scalar ReLU functions.

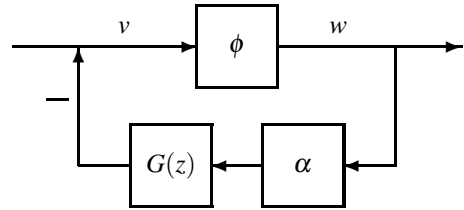
\begin{figure}[h!t]
\centering
\begin{picture}(180,75)(0,-60)
 \thicklines
 \put(50,5){$v$}
 \put(10,0){\vector(1,0){70}}  
 \put(80,-15){\framebox(30,30){$\phi$}}
 \put(110,0){\vector(1,0){70}}  
  \put(130,5){$w$}
 \put(155,0){\line(0,-1){45}}  
 \put(155,-45){\vector(-1,0){20}}  
 \put(105,-60){\framebox(30,30){$\alpha$}}
 \put(105,-45){\vector(-1,0){20}}  
 \put(55,-60){\framebox(30,30){$G(z)$}}
 \put(55,-45){\line(-1,0){20}}  
 \put(35,-45){\vector(0,1){45}}  
 \put(23,-20){\line(1,0){8}}  
\end{picture}
\caption{Lurye System}
\label{fig:Lurye}
\end{figure}

Consider the Lurye system shown in Figure~\ref{fig:Lurye} where $\phi:\R\to\R$ is a nonlinear function, $\alpha$ is a non-negative scaling, and $G(z)=\frac{2z+0.92}{z^2-0.5z}$ is a discrete-time system. The goal is to find the stability margin, i.e. the largest $\alpha\ge 0$ for which this Lurye system is stable for all nonlinearities in this class. 

To obtain the largest value of $\alpha$, we used bisection using the ReLU hard IQCs. The feasibility at each bisection step is a semidefinite program and was solved using CVX \cite{cvx14} as a front-end and SDPT3 \cite{toh99,tutuncu03} as the solver.  The bisection is initialized with $\underline{\alpha}=0$ and $\bar{\alpha}=400$. The bisection is terminated when $\bar{\alpha}-\underline{\alpha} \le 10^{-3} (1+\bar{\alpha})$. This stopping condition implies that the absolute and relative error are both less than $10^{-3}$. 

Table~\ref{tab:LiftingStabResults} shows the stability margins
using several different methods.  The first row shows the results
with our new dynamic IQC for the scalar ReLU
(Lemma~\ref{lem:hardIQCReLU}). The margin improves significantly
with increasing horizon $N$ used in the filter $\Psi_N(z)$.
For comparison, the second row shows stability margins using the method in our previous work \cite{noori24ReLURNN}.  This previous work analyzed stability using a "lifted" plant $G_N$ interconnected
with a repeated ReLU $\Phi:\R^N \to \R^N$.  Our previous stability condition used static QCs for repeated ReLU as specified 
in Lemma~\ref{lem:pointwiseQCs}. The new results with our dynamic IQC (row 1) have significantly larger stability margins than
the previous approach using lifting and static QCs on repeated ReLU (row 2).

The last two rows show similar results but for $[0,1]$ slope
restricted nonlinearities satisfying $\phi(0)=0$.  Specifically, the third row shows the stability margins using dynamic IQCs for this class of nonlinearities (Lemma~\ref{lem:hardIQCSlope}).\footnote{Note that we enforce $P\succeq 0$ in our implementation
as required in Theorem~\ref{thm:slopeRes} to prove internal
stability.  Input-output stability does not require $P$ to
be positive semidefinite. Dropping this constraint, as in
 \cite{carrasco19}, yields almost identical results, for this
example, to those reported in the row 3 of Table~\ref{tab:LiftingStabResults}.} The fourth row uses our
previous approach \cite{noori24ReLURNN}
based on lifting and static QCs for repeated slope restricted nonlinearities
as given in Lemma~\ref{lem:doublyhypQC}. Rows 1 and 2 provide
larger stability margins than the corresponding results in rows
3 and 4.  This confirms that our ReLU QCs/IQCs are supersets of
the corresponding Zames-Falb QCs/IQCs for slope-restricted nonlinearities.  In other words, the larger class of QCs/IQCs for ReLU provide less conservative stability margins.


\begin{table}[h!]
\centering
\scalebox{0.93}{
\begin{tabular}{ |c|c c c c| } 
 \hline
 lift size $N$ & 1 & 2 & 3 & 4\\ \hline
 \textbf{ReLU with dynamic IQCs} & \textbf{0.6504} & \textbf{1.4553} & \textbf{169.6777} & \textbf{221.1914}\\ 
 ReLU with static QCs & 0.6516 & 0.6516 & 1.1734 & 2.2156\\  \hline
 Slope-Res. with dynamic IQCs & 0.6500 & 0.9094 & 0.9109 & 0.9109\\  Slope-Res. with static QCs & 0.6516 & 0.6516 & 0.8072 & 0.8484\\
\hline
 \end{tabular}
 }
\caption{Comparison of stability margins obtained for ReLU and slope-restricted nonlinearities using static QCs and dynamic IQCs with different lifting/filter horizons. Results with our method (row 1) give significantly better margins in this example.}  
    \label{tab:LiftingStabResults}
\end{table}

\section{Conclusions}

This work advanced the stability analysis of discrete-time feedback systems with a ReLU nonlinearity by introducing a class of hard IQCs for scalar ReLU. The proposed construction bridges static QCs and soft IQCs by formulating hard IQCs with Toeplitz-structured matrices that capture the properties of the ReLU nonlinearity. The numerical example demonstrated clear reductions in conservatism relative to Zames–Falb multipliers and earlier static-QC methods, especially as the filter's horizon grows. In future work, we aim to further investigate the relationship between static QC and dynamic IQC methods with the goal of clarifying the connection and superiority between their stability guarantees.

\section{Acknowledgments}

The authors acknowledge AFOSR Grant \#FA9550-23-1-0732 for funding of this work.

\bibliographystyle{myIEEEtran}
\bibliography{references} 
 
\end{document}